\documentclass[seceq]{ptptex}

\usepackage{graphicx}

\title{%        %You can use \\ for explicit line-break
Compton-thick AGN and the Synthesis of the Cosmic X-ray Background: the $Suzaku$ Perspective
}
\author{%       %Use \sc for the family name
Roberto {\sc Gilli}$^{1,}$\footnote{E-mail: roberto.gilli@oabo.inaf.it},
Andrea {\sc Comastri}$^1$,
Cristian {\sc Vignali}$^2$ \\
and G\"unther {\sc Hasinger}$^3$
}

\inst{%         %Affiliation, neglected when [addenda] or [errata]
$^1$INAF - Osservatorio Astronomico di Bologna, Bologna, Italy\\
$^2$Dipartimento di Astronomia, Universit\`a degli Studi di Bologna, Bologna, Italy\\
$^3$Max-Planck-Institut f\"ur extraterrestrische Physik, Garching, Germany\\
}

\recdate{%      %Editorial Office will fill in this.
}

\abst{%       %this abstract is neglected when [addenda] or [errata]
We discuss the abundance of Compton-thick AGN as estimated by the most
recent population synthesis models of the cosmic X-ray
background. Only a small fraction of these elusive objects have been
detected so far, in line with the model expectations. The advances
expected by the broad band detectors on board $Suzaku$ are briefly
reviewed.}

\begin{document}

\maketitle

\section{Introduction}
Despite extensive observational efforts, the population of heavily
obscured, Compton-thick AGN remains elusive, especially at high
redshifts, preventing a complete census of accreting supermassive
black holes (SMBHs). While Compton-thick (CT) nuclei were shown to
hide in about half of local Seyfert 2 galaxies (Risaliti et al. 1999,
Guainazzi et al. 2005), observations of heavily obscured objects
beyond $z\sim0.1$ are very sparse (see Comastri 2004 for a review) and
their abundance can be constrained only by indirect arguments. One
argument is the comparison between the mass function of local SMBHs
with the one expected if they accreted most of their mass during past
AGN phases (Marconi et al. 2004). Another, and probably more
stringent, argument is the residual emission at 30 keV in the spectrum
of the cosmic X-ray background (XRB), which is left after removing the
contribution from the better known population of less obscured,
Compton-thin AGN. The residual 30 keV XRB emission can indeed be
modeled by assuming a population of CT AGN as large as that of
moderately absorbed ones over a broad range of redshifts and
luminosities (see Gilli, Comastri \& Hasinger 2007, hereafter GCH07,
for a recent work). In particular, this residual emission is mostly
filled by ``mildly'' CT objects (defined as those with
$10^{24}<N_H<10^{25}$ cm$^{-2}$) in which the direct, primary emission
is visible above $\sim 10$ keV, rather than by ``heavily'' CT objects
($N_H>10^{25}$ cm$^{-2}$), in which only reflected radiation is
visible at high energy, and are therefore significantly less luminous
than the formers at 30 keV. As a consequence, the number of heavily CT
AGN is poorly constrained even by population synthesis models and is
generally assumed to be similar (equal in GCH07) to that of mildly CT
objects, as suggested by the results of Risaliti et
al. (1999). Because of the strong selection effects due to absorption,
only a small percentage of CT sources have been observed in current
X-ray surveys. In the Chandra Deep Field South (CDFS) only about 5\%
of the detected AGN have been identified as CT candidates (Tozzi et
al. 2006). In the recent $INTEGRAL$/IBIS and $Swift$/BAT surveys
performed above 10 keV, where the absorption bias is less effective, a
higher fraction is observed ($\sim 10-15\%$, Markwardt et al. 2006,
Beckmann et al. 2006). This is already remarkable, if one bears in
mind that X-ray surveys above 10 keV are still limited to very bright
fluxes ($\sim 10^{-11}$ erg cm$^{-2}$ s$^{-1}$). As shown in
Fig.~1~$left$, these small observed fractions are in good agreement
with those expected if CT AGN are intrinsically as abundant as
moderately obscured ones (see GCH07), and are predicted to increase
dramatically at fluxes below the current sensitivity limits.

\begin{figure}[t]
%\epsfxsize=7.cm 
%\epsfbox{Gilli_Roberto_fig1a.ps}
%\epsfxsize=7.cm
%\epsfbox{ngc5728_nufnu.ps}
%\epsfbox{Gilli_Roberto_fig1b.ps}
\includegraphics[angle=0,width=0.48\textwidth]{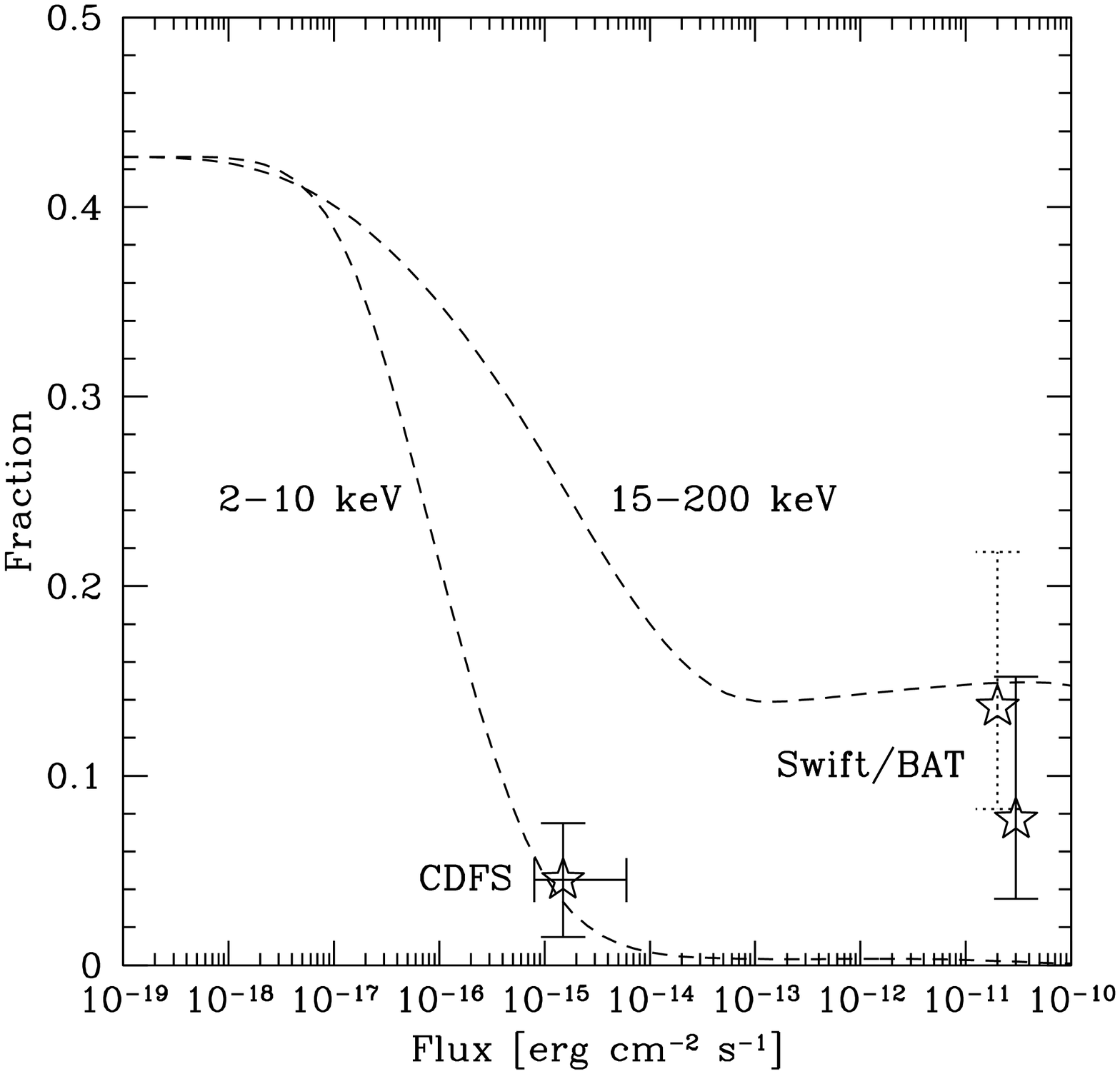}
\hfill
\includegraphics[angle=0,width=0.48\textwidth]{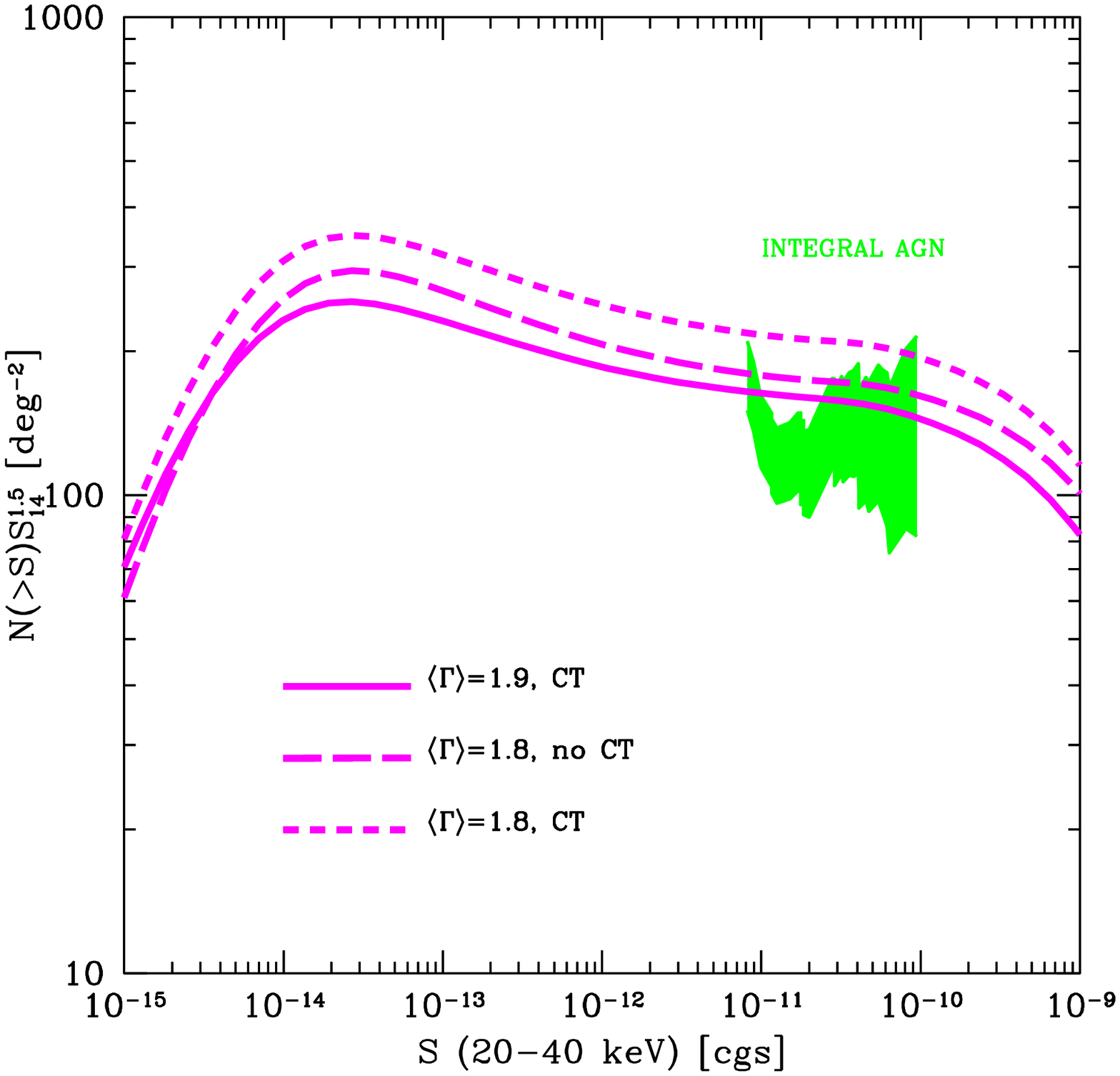}
\caption{$Left:$ The fractions of CT AGN in the GCH07 baseline model
as a function of the 2-10 keV and 15-200 keV limiting fluxes compared
to those observed in the CDFS (Tozzi et al. 2006) and in the
$Swift$/BAT catalog (Markwardt et al. 2005). The upper $Swift$/BAT
point is corrected for incompleteness. Note the steep increase
expected at fluxes below the current sensitivities and the identical
CT fraction in the two bands at extremely faint fluxes, where $all$
AGN should be detected. $Right:$ The predicted 20-40 keV AGN counts
(see Section~2) normalized to an Euclidean Universe and compared with
those measured by $INTEGRAL$ (Beckmann et al. 2006).}
\end{figure}

\begin{figure}[t]
%\epsfxsize=13cm
%\epsfbox{Gilli_Roberto_fig2.ps}
\includegraphics[angle=0,width=0.9\textwidth]{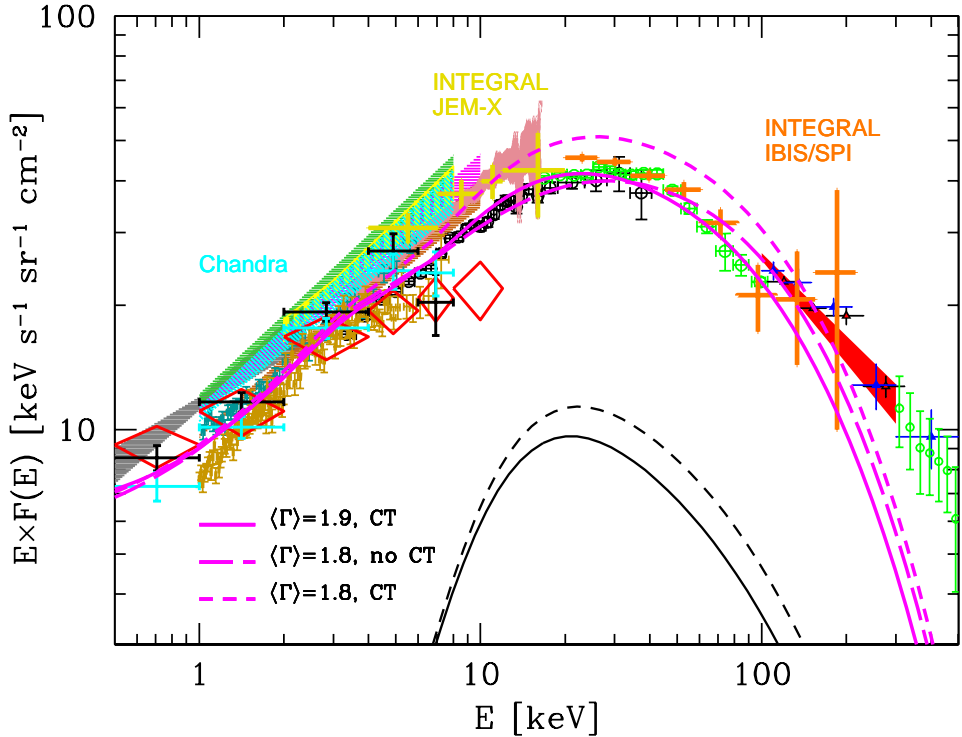}
\caption{The spectrum of the X-ray background. Most of the datapoints
are described in GCH07. Here we add the recent 5-200 keV measurement
by $INTEGRAL$ (Churazov et al. 2007) and the 1-7 keV measurement by
$Chandra$ (Hickox \& Markevitch 2006). Model curves based on GCH07 are
also plotted: the upper magenta curves show the total contribution
from AGN (plus galaxy clusters), according to different assumptions on
the average AGN spectral slope and number of CT objects as labeled
(see text); the lower black curves represent the corresponding CT
contribution.}
\end{figure}

\section{Uncertainties on the number of Compton-thick AGN}

Since the overall abundance of CT AGN is estimated by subtracting from
the XRB spectrum the contribution of Compton-thin sources, it is
imperative to estimate the latter at the best of present
knowledge. The modeling presented in GCH07 took into account a
detailed characterization of the average AGN X-ray spectra, including
dispersion, and cosmological evolution, but is nonetheless worth
exploring the parameter space to some extent and check how different
assumptions may affect the estimated CT number. In the baseline model
presented by GCH07, a Gaussian distribution in the AGN primary
continuum was considered, with average spectral slope and dispersion
of $\langle\Gamma\rangle=1.9$ and $\sigma_{\Gamma}=0.2$, respectively,
in agreement with the observed distributions (Mateos et al. 2005). In
Fig.~2 we show the effects of assuming an average spectral index
$\langle\Gamma\rangle=1.8$ with the same dispersion. A sligthly harder
average spectral powerlaw is in principle sufficient to saturate the
XRB emission with Compton-thin AGN, leaving little room for CT
sources. Indeed, when adding as many CT AGN as in the baseline model,
the 30 keV XRB emission measured by HEAO-1, and recently confirmed
within 10\% by $INTEGRAL$ (Churazov et al. 2007), is exceeded if
$\langle\Gamma\rangle=1.8$. Furthermore, the baseline model appears to
be in much better agreement with other observational constraints, such
as the spectral distributions observed in different AGN samples and
the observed numbers of CT AGN (see GCH07). The model predictions have
been further compared with the AGN counts in the 20-40 keV band
recently estimated by Beckmann et al. (2006). The situation
(Fig.~1~$right$) is similar to that shown in Fig.2 for the XRB,
although the constraints are less stringent. While a model with
$\langle\Gamma\rangle=1.8$ would imply a small number of CT AGN, the
baseline model provides a good match to the data with a relative ratio
of one between Compton-thick and Compton-thin AGN at all
redshifts. This assumed ratio appears more in line with current
observations both in the local (Risaliti et al. 1999) and in the
distant Universe (Martinez-Sansigre et al. 2006).

\section{The $Suzaku$ perspective}

To date, about 40 local AGN have been shown to be CT through X-ray
observations (Comastri 2004) and their number is expected to increase
significantly in the next future. Indeed, CT AGN candidates in current
$INTEGRAL$ and $Swift$ surveys can be easily flagged as such if their
X-ray flux above 10 keV is much larger than their soft X-ray flux,
which is often available from archival X-ray data. Like $BeppoSAX$ in
the past years, $Suzaku$ is now carrying on board detectors which are
sensitive in the broad 0.5-50 keV band and are therefore the ideal
instruments to determine the X-ray spectral energy distribution of
bright, nearby CT objects. In EAO-1 we have obtained $Suzaku$
observations of 2 CT candidates selected from the $INTEGRAL$ and
$Swift$ AGN catalogs (two further candidates will be observed in
EAO-2). One of the two objects indeed proved to be CT, while the other
turned out to be heavily obscured but still Compton thin (see Comastri
et al. this volume, for a more detailed discussion). $Suzaku$
observations of additional CT candidates selected above 10 keV are
being performed by other groups (e.g. Ueda et al., this
volume). Eventually, once these programs are put together to get
sufficient object statistics, the fraction of CT AGN in the local
Universe will be determined to better accuracy. In particular, new
mildly CT AGN should be revealed in significant numbers, and a few
spectra of heavily CT AGN, which went undetected by $BeppoSAX$, may be
also obtained.

\section*{Acknowledgements}

We are grateful to Mike Revnivtsev and Volker Beckmann for providing
the $INTEGRAL$ XRB spectrum and 20-40 keV AGN counts,
respectively. RG, AC and CV acknowledge financial support from the
Italian Space Agency (ASI) under the contract ASI--INAF I/023/05/0.

\end{document}